\documentclass[a4paper,superscriptaddress,onecolumn,showkeys]{revtex4-2}
\usepackage{bm}
\usepackage{graphicx}
\PassOptionsToPackage{version=3}{mhchem}
\usepackage{mhchem}
\usepackage{xcolor}
\usepackage[italicdiff]{physics}

\begin{document}
\title{Twisted light-induced spin-spin interaction in a chiral helimagnet}
\author{Yutaro Goto}
\email{goto-2@pe.osakafu-u.ac.jp}

\affiliation{Department of Physics and Electronics, Osaka Prefecture University,
Osaka 599-8531, Japan}
\author{Hajime Ishihara}
\affiliation{Department of Physics and Electronics, Osaka Prefecture University,
Osaka 599-8531, Japan}
\affiliation{Department of Materials Engineering Science, Osaka University, Osaka
560-8531, Japan}
\affiliation{Quantum Information and Quantum Biology Division, Institute for Open
and Transdisciplinary Research Initiatives, Osaka University, Osaka
560-8531, Japan}
\author{Nobuhiko Yokoshi}
\email{yokoshi@pe.osakafu-u.ac.jp}

\affiliation{Department of Physics and Electronics, Osaka Prefecture University,
Osaka 599-8531, Japan}
\begin{abstract}
We theoretically investigate how the orbital angular momentum of light can affect a chiral magnetic order. Here, we consider a metallic chiral helimagnet, which is under stationary radiation of a resonant optical vortex beam. We propose a novel interaction between local spins considering microscopic interactions between an optical vortex and electrons.  This vortex-induced interaction modulates the chiral magnetic order in an entirely different way than an external magnetic field does.  Our spin modulation technique may pave a route to create a unique topological or chiral structure for future opto-spintronics devices.
\end{abstract}
%
\maketitle

\section{Introduction}
Chiral helimagnets are ferromagnetic materials belonging to chiral space groups or non-centrosymmetric space groups. The local spin moments of such magnets are located on the chiral framework of the lattice; thus, its magnetization is twisted. Chiral helimagnetic materials have promising spintronic functionalities for device applications. This is because they have various physical properties owing to nonlinearity,
robustness, topology, and tunability \citep{RN359}. In chiral helimagnet, the chiral spin texture, which means the left- or right-handed helical spin texture along the crystallographic axis, is characterized by the antisymmetric Dzyaloshinskii-Moriya (DM) interaction specified by
the DM vector $\bm{D}$ along the helical axis \citep{RN170,RN540} and Heisenberg interaction specified by $J$.

On the other hand, in the aspect of chirality, Allen \textit{et al.} proposed the twisted light with orbital angular momentum (OAM), so-called optical vortex \citep{RN468}. The optical vortex carries its intrinsic
OAM and has the helical phase structure and toroidal field intensity resulting from the topological singularity along the propagating axis. Recently, many applications of the optical vortex are proposed in broad fields, e.g., stimulated emission depletion microscopy \citep{RN471}, mechanical rotation control of classical micro-particles \citep{RN469,RN3},
laser ablation using a chiral beam \citep{RN117}, and shaping some complex shapes of microfibers \citep{RN494}. Besides, as some attempts to transfer optical OAM to magnetic materials, Fujita \textit{et al.} proposed an ultrafast generation of skyrmionic defects due to a heat distribution of a vortex beam and a creation of anisotropic spin waves and topological defects in chiral magnets due to a magnetic field of the beam \citep{RN220,RN450}.

Theories in magneto-optics since Faraday's era say that the optical response of magnets stems from transferring the angular momentum of light to that of electrons through the optical transition. Conventionally in this situation, optical angular momentum mostly refers only to spin angular momentum (SAM), i.e., the circular polarization helicity. In analogy with transferring the optical SAM to electrons, the optical
OAM can also be transferred via the spin-orbit interaction. Some theoretical and experimental studies show that optical transitions by the vortex are different from those of traditional plane waves \citep{RN358,RN354,RN533}.

We focus on a one-dimensional (1-D) chiral helimagnet, e.g., chromium-intercalated niobium disulfide ($\ce{CrNb3S6}$), classified as a metal where itinerant and localized electrons coexist \citep{RN359}. Various theoretical and experimental works have revealed that the competition between Heisenberg interaction and DM interaction characterizes the chiral magnetic structure \citep{RN493,RN509,RN524,RN109,RN229}. The DM interaction in this material acts between the localized spins mediated by the itinerant electrons. This reminds us of the existence of Ruderman--Kittel--Kasuya--Yosida (RKKY) interaction. Although the RKKY interaction is usually invariant for spin rotation, the coexistence of the spin-orbit interaction and the spatially distributed magnetic moments gives rise to the anisotropic effects \citep{RN610,RN611}. In this magnet, the local magnetic moments originating from the OAM of the localized electron in Cr sites polarize the itinerant electrons locally through the spin-orbit interaction. The RKKY picture, where the itinerant electrons are scattered by the local magnetic moment, is sometimes called the generalized RKKY picture \citep{RN609}. The RKKY interaction with the anisotropic effect causes the DM interaction in this material \citep{RN359}. The optical vortex and the chiral helimagnetic order have a similar spatial structure from the viewpoint of chirality; a kind of chiral couplings between them can be expected. The RKKY process through electron hopping under the optical vortex radiation can affect the interaction between nearest-neighbor local spins nontrivially. We show that this optical vortex-induced spin-spin interaction can modulate the chiral helimagnetic order of $\ce{CrNb3S6}$. The interaction creates the possibility of microscopically modulating the axis of rotation of the helical structure, giving the macroscopic helical magnetic structure a new degree of freedom. Such modulation cannot be achieved by the Zeeman effect and magnetic resonance. There are various proposals to control magnetic interactions with electromagnetic fields \citep{RN37,RN548,RN612,RN613,RN614,RN615}. In addition to these methods, the control of spin-spin interaction by the OAM of light can be a new avenue for the development of magneto-optics or opto-spintronics.

\section{Model}
\begin{figure}
\begin{centering}
\includegraphics[width=0.65\linewidth]{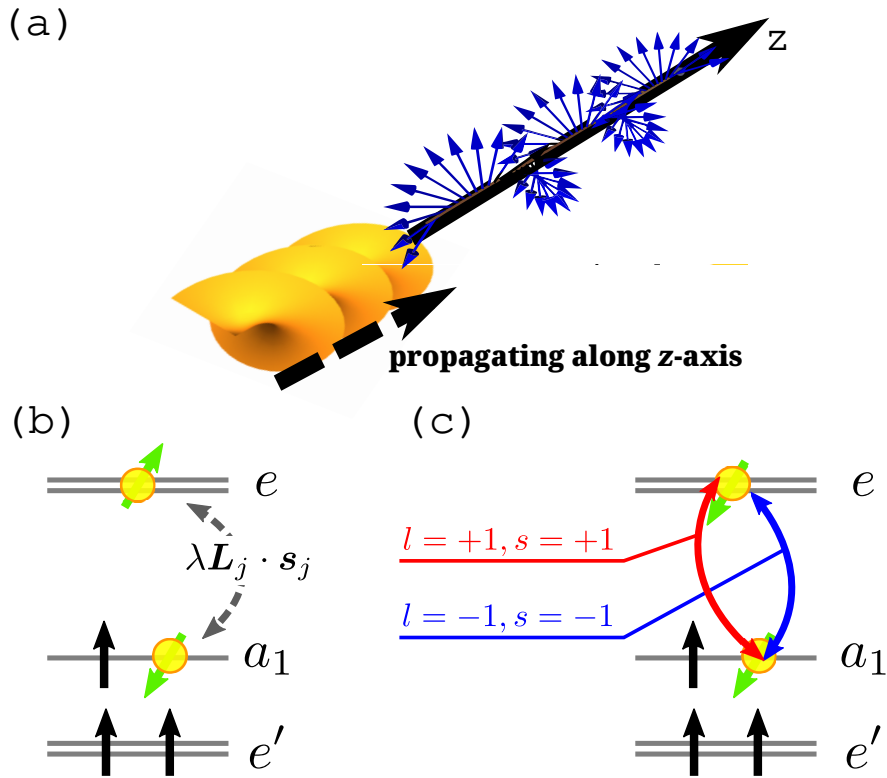}
\par\end{centering}
\caption{(a) Schematic image of the system setup. The optical vortex propagates toward the helical axis ($z$-axis) of the 1-D chiral helimagnet. The DM vector $\bm{D}$ is directed to the positive $z$-axis and that the incident LG beam propagates to the $z$-direction as a circularly polarized light. (b) Schematic of the electronic structure and electronic transition under consideration. The $d$-orbitals in $\ce{CrNb3S6}$ with $D_{3}$ symmetry split into a nondegenerate level $a_{1}$ and two double degenerate levels $e$ and $e^{\prime}$ by the crystalline field. Black and green solid arrow refer to local and itinerant spins, respectively. The spin-orbit coupling $\bm{L}_{j}\cdot\bm{s}_{j}$ excites the itinerant electron from the state $e$ to the state $a_{1}$ with spin-flipping. (c) The optical vortices with OAM $\left(l\right)$
and SAM $\left(s\right)$: $\left(l,s\right)=\left(\pm1,\pm1\right)$
allow the inner d-orbital optical transition without spin-flipping between orbital $a_{1}$ and $e$ as shown by the red and blue arrows.}
\label{fig:LightAbsorbtion}
\end{figure}

We model the optical vortex beam by the Laguerre-Gaussian (LG) mode. Here, we assume that the helical magnetic structure and the LG beam are coaxial (see figure \ref{fig:LightAbsorbtion}(a)). The LG beam is represented by the four-vector $A^{\mu}=\left(\varphi/c,\bm{A}\right)$ satisfying the Helmholtz equation. Adopting Lorenz gauge $\partial_{\mu}A^{\mu}=0$, we obtain a traveling wave in the $z$-direction with the frequency $\omega$ and the wave number $k$. It becomes within the paraxial approximation
\begin{align}
\varphi^{\left(l,s\right)}(\rho,\eta,z,t)= & \frac{c^{2}}{i\omega}\nabla\cdot\bm{A}^{\left(l,s\right)}(\rho,\eta,z,t),\\
\bm{A}^{\left(l,s\right)}(\rho,\eta,z,t)= & \bm{A}_{0}^{s}\sqrt{\frac{2}{\pi\abs{l}!}}\frac{1}{w\left(z\right)}\left[\frac{\rho\sqrt{2}}{w\left(z\right)}\right]^{|l|}\exp\left[-\frac{\rho^{2}}{w\left(z\right)^{2}}\right]\exp\left[i\left(\frac{k\rho^{2}}{2R\left(z\right)}+l\eta-\psi\left(z\right)\right)\right]e^{i\left(kz-\omega t\right)}, \label{eq:vector potential}
\end{align}
where we use the cylindrical coordinate $\left(\rho,\eta,z\right)=\left(\sqrt{x^{2}+y^{2}},\tan^{-1}\left(y/x\right),z\right)$,
and $\bm{A}_{0}^{s}/|\bm{A}_{0}^{s}|$ represents the polarization of light. Here, $\left(l,s\right)$ indicates the OAM and SAM of light, $w\left(z\right)$ is the beam radius, $R\left(z\right)$ the radius of curvature of the beam and $\psi\left(z\right)$ the Gouy phase.

In describing the electronic system in the chiral helimagnet, we treat the itinerant electrons in a multi-orbit tight-binding model, in which they interact with the localized spins via strong Hund's coupling;
\begin{align}
\mathcal{H}_{\text{ch}} &
=\mathcal{H}_{0}+\mathcal{H}_{\text{hop}}+\mathcal{H}_{\text{so}}
=\sum_{i,\mu,\gamma}E_{\mu}c_{i\mu\gamma}^{\dagger}c_{i\mu\gamma}-J_{\text{H}}\sum_{i}\bm{S}_{i}\cdot\bm{s}_{i}+\mathcal{H}_{\text{hop}}+\mathcal{H}_{\text{so}},\label{eq:Hubbard original}
\end{align}
where $E_{\mu}$ is the energy of $d$-electrons. The field operator of the electron is $\psi_{\mu\gamma}({\bm{r}})=\Sigma_{i}\phi_{i\mu}({\bm{r}})c_{i\mu\gamma}$ where $\phi_{i\mu}({\bm{r}})$ are the Wannier function centered on the site ${\bm{R}}_{i}$ with $c_{i\mu\gamma}$ being the corresponding annihilation operator of the itinerant electrons. The indices $\mu$ and $\gamma$ indicate the orbit and spin. The constant $J_{\text{H}}$ denotes the Hund's coupling energy between the itinerant and localized electron spins. We define the itinerant spin as $\bm{s}_{i}=(1/2)\Sigma_{\mu,\gamma,\gamma^{\prime}}c^{\dagger}_{i\mu\gamma} \langle \gamma| \bm{\sigma} |\gamma^{\prime}\rangle c_{i\mu\gamma^{\prime}}$ with $\bm{\sigma}$ being Pauli matrix, while the operator $\bm{S}_{i}$ is the localized spin. The hopping and the spin-orbit interaction are
\begin{align}
\mathcal{H}_{\text{hop}} & =-\Gamma \sum_{i,j,\mu,\nu,\gamma}c_{i\mu\gamma}^{\dagger}c_{j\nu\gamma},\\
\mathcal{H}_{\text{so}} & =\lambda\sum_{i}\bm{L}_{i}\cdot\bm{s}_{i},\label{eq:Spin-orbitInteraction}
\end{align}
where $\Gamma$ and $\lambda$ are the corresponding coupling constants, and $\bm{L}_{i}$ is the OAM of the ion on the site ${\bm{R}}_{i}$ \citep{RN170,RN540}. Figure \ref{fig:LightAbsorbtion} shows the schematic of the electronic level structure and the electronic transition by the spin-orbit interaction and optical vortex absorption. We consider only the spin-orbit coupling $\bm{L}_{i}\cdot\bm{s}_{i}$ between the ion and itinerant spin. It contributes to excite the itinerant electron from a higher state $e$ to a lower state $a_{1}$ with spin-flipping (figure \ref{fig:LightAbsorbtion}(b)). The site-dependent $\bm{L}_{i}$ corresponds to the local magnetic moment, which reflects the crystal symmetry of the helimagnet. We have neglected the spin-orbit interaction between the local momenta $\bm{L}_{i}\cdot\bm{S}_{i}$. It contributes to the interband transition with spin-flipping whose transition energy is much larger than that by $\bm{L}_{i}\cdot\bm{s}_{i}$ within the Hund's rule. In addition, considering the effect of the optical vortex absorption, the $\bm{L}_{i}\cdot\bm{S}_{i}$ term should be higher-order process in the RKKY scheme.

The 1-D helimagnetic material $\ce{CrNb3S6}$ is the transition metal dichalcogenide ($\ce{NbS2}$) intercalated with $3d$ magnetic elements ($\ce{Cr}$). The localized spin with $S=|\bm{S}_{i}|=3/2$ is originated from the localized electron in Cr atoms in trivalent state. On the other hand, the itinerant $d$-electrons belong to the $\ce{NbS2}$-derived unfilled band. As the $\ce{Cr}$ atom plays a role of the bridge between the well-separated $\ce{NbS2}$ layers, the interlayer electronic conduction occurs via the $\ce{Cr}$ atom. Thus, the itinerant and localized electrons couple with each other. The local symmetry around $\ce{Cr}$ is $D_{3}$ and its $3d$-orbitals split into $e,a_{1}$ and $e^{\prime}$ orbitals \citep{RN38}. Here as the orbital $e$ has $d_{x^{2}-y^{2}}/d_{xy}$ symmetries and the orbital $a_{1}$ has $d_{3z^{2}-r^{2}}$ symmetry, the interband transition of the d-electron by optical vortices via the dipole interaction is allowed for $\left(l,s\right)=\left(\pm1,\pm1\right)$ (see figure \ref{fig:LightAbsorbtion}(c)) \citep{RN306}. Note that such an elementary process is forbidden for plane waves with linearly or circularly polarized light due to the transition selection rule, and requires the OAM of light. Thus, we sum over $a_{1}$ and $e$ for the orbitals $\mu$ and $\nu$ in the above Hamiltonian.

The `vector potential' in the presence of the spin-orbit interaction is defined by $\bm{\chi}^{\left(l,s\right)}\left(\bm{r},t\right)=-e\bm{A}^{\left(l,s\right)}\left(\bm{r},t\right)-\left(\mu_{\text{B}}/2c^{2}\right)\bm{\sigma}\times\bm{E}^{\left(l,s\right)}\left(\bm{r},t\right)$ \citep{RN29}. Here, $\mu_{\text{B}}$ is the Bohr magneton, and $\bm{E}^{\left(l,s\right)}\equiv i\omega\bm{A}^{\left(l,s\right)}-\nabla\varphi^{\left(l,s\right)}$. We introduce the vector potential through the on-site interband excitation , i.e., $\mathcal{H}^{\prime} =\mathcal{H}_{\text{ch}}+\mathcal{H}^{\bm{\chi}}$, where
\begin{align}
\mathcal{H}^{\bm{\chi}}(t)& =\sum_{i,\mu\neq\nu,\gamma,\gamma^{\prime}}\bra{i\mu\gamma}\frac{1}{2m}\left(\bm{p}\cdot\bm{\chi}^{\left(l,s\right)}\left(\bm{r},t\right)+\bm{\chi}^{\left(l,s\right)}\left(\bm{r},t\right)\cdot\bm{p}+\text{H.c.}\right)\ket{i\nu\gamma^{\prime}}c_{i\mu\gamma}^{\dagger}c_{i\nu\gamma^{\prime}}
\label{el-vortex-int}
\end{align}
where $m$ and $\bm{p}$ are the mass and momentum of electron. In addition, we have defined the matrix element for some operator $\bm{\mathcal{O}}$ as $\mel{i\mu\gamma}{\bm{\mathcal{O}}}{j\nu\gamma^{\prime}}=\mel{\gamma}{\int d\bm{r}\phi_{i\mu}^{*}\left(\bm{r}\right)\bm{\mathcal{O}}\left(\bm{r}\right)\phi_{j\nu}\left(\bm{r}\right)}{\gamma^{\prime}}$. The form of equation (\ref{el-vortex-int}) itself does not assume a specific electric field. For instance, when $\bm{E}^{\left(l,s\right)}\left(\bm{r},t\right)$ is constant in time and space, it reduces to the spin-orbit interaction of Bychkov-Rashba type \citep{RN608}. Here we consider the oscillating electric field of the optical vortex. The energy of the electric field is resonant with the interband transition, i.e., $\hbar \omega=E_{\text{g}}$ where $E_{\text{g}}=E_{e}-E_{a_{1}}+SJ_{\text{H}}$ is the gap energy including the Hund's coupling. Thus, one can consider that the itinerant electrons have two spin-orbit interactions, one with the localized moment $\bm{L}_{i}$ at each site and the other with the optical vortex in this system.

\section{Effective 1-D Spin Model}
We derive the effective model considering the perturbations with respect to the electron hopping $\mathcal{H}_{\text{hop}}$, spin-orbit interaction $\mathcal{H}_{\text{SO}}$, and optical vortex absorption $\mathcal{H}^{\bm{\chi}}$. The unperturbed Hamiltonian is $\mathcal{H}_0$, and the itinerant electrons in the lower band ($a_1$) have their spins parallel to the localized spins in the unperturbed ground state. Thus, we impose the constraint not to break the Hund rule throughout the perturbation calculation \citep{RN174}. Here we consider the time evolution operator in interaction picture
\begin{align}
U(t,0)=\mathcal{T}
\left\{
\exp  \left[-\frac{i}{\hbar}\int_0^t (\mathcal{H}_{\text{hop}}+\mathcal{H}_{\text{SO}}+\mathcal{H}^{\bm{\chi}}(t^{\prime})) dt^{\prime}
\right]
\right\},
\end{align}
where $\mathcal{T}$ is the time-ordering operator. We expand this operator for each order of the perturbation $U(\eta,0)=\Sigma_{n\ge 0}U^{(n)}(t,0)$. The Hamiltonian from $n$-th perturbation can be written as
\begin{align}
\mathcal{H}^{(n)}(t)=
\sum_{i,j,\gamma,\gamma^{\prime}}
\left\{
i\frac{d}{dt}\langle i, a_1, \gamma|U^{(n)}(t,0)|j, a_1, \gamma^{\prime} \rangle
\right\}
c_{ia_1\gamma}^{\dagger}
c_{ja_1\gamma^{\prime}}.
\end{align}
Throughout the perturbation calculation, we use a so-called secular approximation to neglect components with the fast oscillation \citep{RN616,RN617}. Thus, we can consider only the time-independent Hamiltonian. This approximation is safely justified because the perturbation energies are much smaller than the gap energy $E_{\text{g}}$. In addition, the period of the neglected oscillation $\sim 2\pi \hbar/E_g$ is much shorter than the time step of the subsequent simulation. Such an oscillation should be averaged out during the time step.

In the first-order perturbation, only the hopping Hamiltonian $\mathcal{H}_{\text{hop}}$ is relevant. Due to the Hund rule, the localized spins tend to be aligned for the itinerant electrons to have more extended wave forms. Thus, the first-order Hamiltonian makes the localized spins exhibit ferromagnetic tendency, and the exchange coupling $J$ between the localized spins is proportional to the hopping energy $\Gamma$ \citep{RN416}. The details of the perturbation calculations are described in the Supplementary Material. In the second-order perturbation, the products of the spin-orbit interaction and the hopping energy remain. In the third-order perturbation, the triple product of the hopping energy, the spin-orbit interaction, and optical vortex absorption is relevant. By using the method of projection operator, the results of the perturbation calculation can be summarized as follows \citep{RN416},
\begin{align}
\mathcal{H}^{\prime}  & =\mathcal{H}_{0}-\Gamma \sum_{i,j,\mu,\nu,\gamma,\gamma^{\prime}}c_{i\mu\gamma}^{\dagger}\bra{\gamma}P_{i}P_{j}\ket{\gamma^{\prime}}c_{j\nu\gamma^{\prime}}.
\label{eq:HamiltonianPerturbated}
\end{align}
The second term represents the complex one-particle hopping process, which incorporates the effects due to the spin-orbit interaction and the optical vortex absorption. The projection operator at the $i$-th site is
\begin{align}
P_{i} & =\frac{\left(\bm{S}_{i}+\bm{v}_{i}+\bm{v}_{i}^{\prime}\right)\cdot\bm{\sigma}+\left(S+1\right)}{2S+1},
\end{align}
where
\begin{align}
\bm{v}_{i} & =-\frac{i\lambda}{E_{g}}\mel{a_{1}}{\bm{L}_{i}}{e}\times\bm{S}_{i},\\
\bm{v}_{i}^{\prime} & =\frac{2i\lambda}{\hbar\delta} \mel{a_{1}}{\bm{L}_{i}}{e}\times\left(\Im\mel{a_{1}}{\bm{B}_{i}^{\left(l,s\right)}}{e}\times\bm{S}_{i}\right).
\label{eq:vprime}
\end{align}
with $\delta$ being the phenomenological line width for the optical transition, and $\bm{B}_{i}^{\left(l,s\right)}=(-\mu_{\text{B}}/2c^{2})\bm{E}_{i}^{\left(l,s\right)}\times\bm{r}_{i}$. As can be seen from equation (\ref{eq:vprime}), the absorption line width $\delta$ is essential for the third-order perturbation to account for the resonant absorption of the optical vortex. This line width dampens some of the oscillations and allows us to define a time-independent perturbation Hamiltonian at the steady state where $\exp(-\delta t)\sim 0$ (see the Supplementary Material). In the absence of the width ($\delta=0$), the perturbation calculation is violated and the time-independent perturbative Hamiltonian is not defined. In addition, we ignore the perturbation terms that contain more than two optical vortex absorption terms. This is because such terms should be small in the parameter regime considered in this study. The three vectors $\bm{S}_{i}$, $\bm{v}_{i}$, and $\bm{v}_{i}^{\prime}$ incorporate the effects of the first-, the second-, and the third-order perturbation, respectively. We can obtain the effective Hamiltonian for the localized spins  by making the wavefunction of the itinerant electron extend the most in equation \eqref{eq:HamiltonianPerturbated} \citep{RN416}.

Mapping the effective Hamiltonian onto  1-D spin model, we obtain the Hamiltonian for the localized ion spin;
\begin{align}
\frac{\mathcal{H}_{\text{1D}}}{S^{2}} & =-J\sum_{i}\hat{\bm{n}}_{i}\cdot\hat{\bm{n}}_{i+1}+\sum_{i}\bm{D}_{i}\cdot\left(\hat{\bm{n}}_{i}\times\hat{\bm{n}}_{i+1}\right)-\sum_{i}\hat{\bm{n}}_{i}\cdot\left(\bm{D}_{i}\bm{\mathcal{B}}_{i}^{\left(l,s\right)}+{\bm{\mathcal{B}}}_{i}^{\left(l,s\right)}\bm{D}_{i}\right)\cdot\hat{\bm{n}}_{i+1},\label{eq:EffectiveHamiltonian2}
\end{align}
where $J\sim \Gamma/S^{2}$ is the Heisenberg exchange coupling, and
\begin{align}
 & \bm{D}_{i}=-\frac{\lambda J}{E_{\text{g}}}\Im\left(\mel{a_{1}}{\bm{L}_{i}}{e}-\mel{a_{1}}{\bm{L}_{i+1}}{e}\right),\label{eq:DMvector}\\
 & \bm{\mathcal{B}}_{i}^{\left(l,s\right)}=\frac{2E_{\text{g}}}{\hbar\delta}\Im\left(\mel{a_{1}}{\bm{B}_{i}^{\left(l,s\right)}}{e}-\mel{a_{1}}{\bm{B}_{i+1}^{\left(l,s\right)}}{e}\right).\label{eq:B}
\end{align}Here $\hat{\bm{n}_{i}}$ are the magnitude and the direction
of the localized ion spins ($\bm{S}_{i}=S\hat{\bm{n}}_{i}$). The DM interaction vector between the adjacent spins $\bm{D}_{i}$ originates in the vector $\bm{v}_{i}$. The RKKY interaction usually depends on the Fermi wavelength. In the present calculation, instead of describing the mediation of conduction electrons by scattering of wave functions, we use a hopping Hamiltonian. In addition, the energies of the initial and intermediate states are described by the band energy without considering the wavenumber dependence. Thus, the Fermi wave number does not appear in the DM interaction vector. One can consider the differences in local OAMs between adjacent sites are the same everywhere, i.e., $\bm{D}_{i}=\bm{D}$. When the optical vortex-induced term is absent, equation \eqref{eq:EffectiveHamiltonian2}
is known to well explain various magnetic features of $\ce{CrNb3S6}$ with $\ce{Cr}$ sites twisting \citep{RN170,RN540,RN359}. In other words, we effectively take into account the helical structure of the $\ce{Cr}$'s spins through equation \eqref{eq:DMvector}.

\begin{figure}
\begin{centering}
\includegraphics[width=0.7\linewidth]{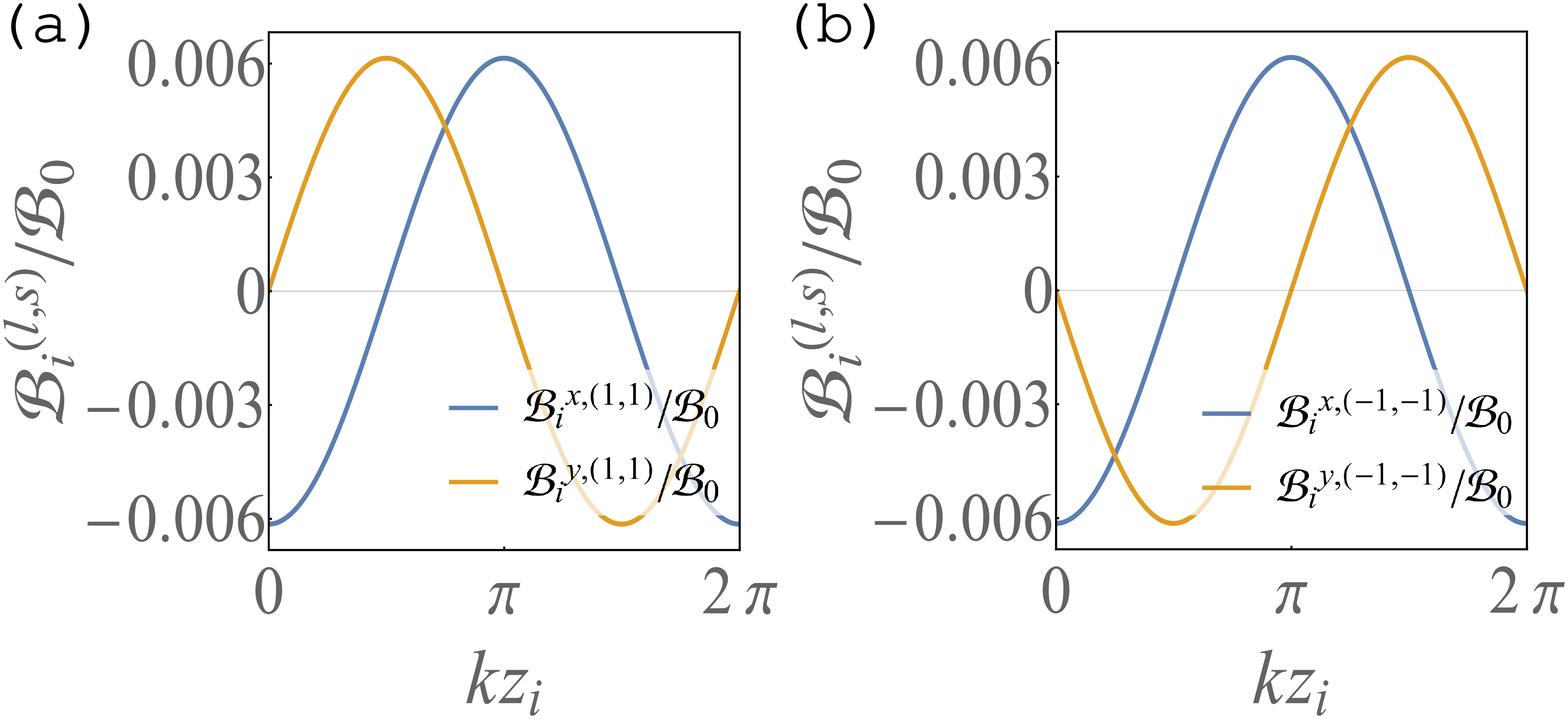}
\par\end{centering}
\caption{The spatial profile of the intensity of the vortex-spin
coupling $\mathcal{B}_{i}^{\left(l,s\right)}\left(z_{i}\right)/\mathcal{B}_{0}$.
The vector $\bm{\mathcal{B}}_{i}^{\left(l,s\right)}/\mathcal{B}_{0}=\sin\left(ka/2\right) \bm{b}_{i}^{\left(l,s\right)}$ oscillates with $z$  in the $x$-$y$ plane since $\bm{b}_{i}^{\left(\pm1,\pm1\right)}=-\cos \left[ k \left( z_{i}+a/2\right) \right] \hat{\bm{e}}^{x}\pm\sin \left[ k \left( z_{i}+a/2\right) \right]\hat{\bm{e}}^{y}$ with quantum numbers (b) $\left(l,s\right)=\left(1,1\right)$ and (c) $\left(-1,-1\right)$.
The $x$- and $y$-component are shown as the blue and orange lines, respectively.}
\label{fig:SchematicImage}
\end{figure}

The last term in the right-hand side of equation \eqref{eq:EffectiveHamiltonian2}
is the optical vortex-induced spin-spin interaction and originates in the vector $\bm{v}_{i}^{\prime}$. Note that the vector $\bm{\mathcal{B}}_{i}^{\left(l,s\right)}$
does not refer to the magnetic field of light. It is proportional
to the LG beam intensity and is the dimensionless vector that determines the coupling between the LG beam and the spins.
We assume that the DM vector points in the positive direction of $z$-axis ($\bm{D}=D\hat{\bm{e}}^{z}$,
$D>0$) with $\hat{\bm{e}}^{\xi}$ being the $\xi$-directed unit vector. As for the optical vortex, we assume the LG beam with $(l,s)=(\pm 1,\pm 1)$ and focus on the light-induced effect near the focal spot. We neglect the $z$-dependence of $w\left(z\right)$, $R\left(z\right)$, and $\psi\left(z\right)$ in the LG beam.  In this case, the vector $\bm{\mathcal{B}}_{i}^{\left(l,s\right)}=\mathcal{B}_{0}\sin\left(ka/2\right) \bm{b}_{i}^{\left(l,s\right)}$ oscillates with $z$  in the $x$-$y$ plane since $\bm{b}_{i}^{\left(\pm1,\pm1\right)}=-\cos \left[ k \left( z_{i}+a/2\right) \right] \hat{\bm{e}}^{x}\pm\sin \left[ k  \left( z_{i}+a/2\right) \right]\hat{\bm{e}}^{y}$ (see figure \ref{fig:SchematicImage}). Here,  $a=z_{i+1}-z_{i}$ is the lattice constant. We rewrite the last term as follows;
\begin{align*}
 -2D\mathcal{B}_0 \sin \left( \frac{ka}{2} \right) \sum_{i} {\bm{n}}_{i}\cdot\left(\hat{\bm{e}}_{i}^{+} \hat{\bm{e}}_{i}^{+}
 -\hat{\bm{e}}_{i}^{-} \hat{\bm{e}}_{i}^{-}
 \right) \cdot{\bm{n}}_{i+1},
\end{align*}
where $\hat{\bm{e}}_{i}^{\pm}=(\hat{{\bm{b}}}_{i}^{\left(l,s\right)}\pm\hat{\bm{e}}^{z})/\sqrt{2}$.
Since the Heisenberg coupling is dominant in this system, this term
tends to align the $\hat{\bm{e}}_{i}^{+}$-components of the adjacent
local spins. On the contrary, for the chiral magnet with $D<0$, the term tends to align the $\hat{\bm{e}}_{i}^{-}$-components. In this sense, one can interpret this vortex-induced interaction as a kind
of chiral couplings between the magnetic structure and the vortex beam.

\section{Steady State Magnetic Structure}
In order to investigate the effect of the vortex-induced interaction,
we numerically determine the steady-state magnetic order of the 1-D
chiral helimagnet. By defining the normalized effective magnetic field
as $\hat{\bm{h}}_{i}\equiv-1/JS^{2}\left(\delta\langle\mathcal{H}_{\text{1D}}\rangle/\delta\hat{\bm{n}}_{i}\right)$,
the Landau--Lifshitz--Gilbert (LLG) equation becomes
\begin{equation}
\frac{d\hat{\bm{n}}_{i}}{d\tau}=-\frac{1}{1+\alpha^{2}}\left[\hat{\bm{n}}_{i}\times\hat{\bm{h}}_{i}+\alpha\hat{\bm{n}}_{i}\times(\hat{\bm{n}}_{i}\times\hat{\bm{h}}_{i})\right],\label{eq:LLG}
\end{equation}
where $\alpha$ is the Gilbert damping constant. We use the dimensionless time $\tau$ normalized by $\hbar/(JS)$.

\begin{figure}
\begin{centering}
\includegraphics[width=0.85\linewidth]{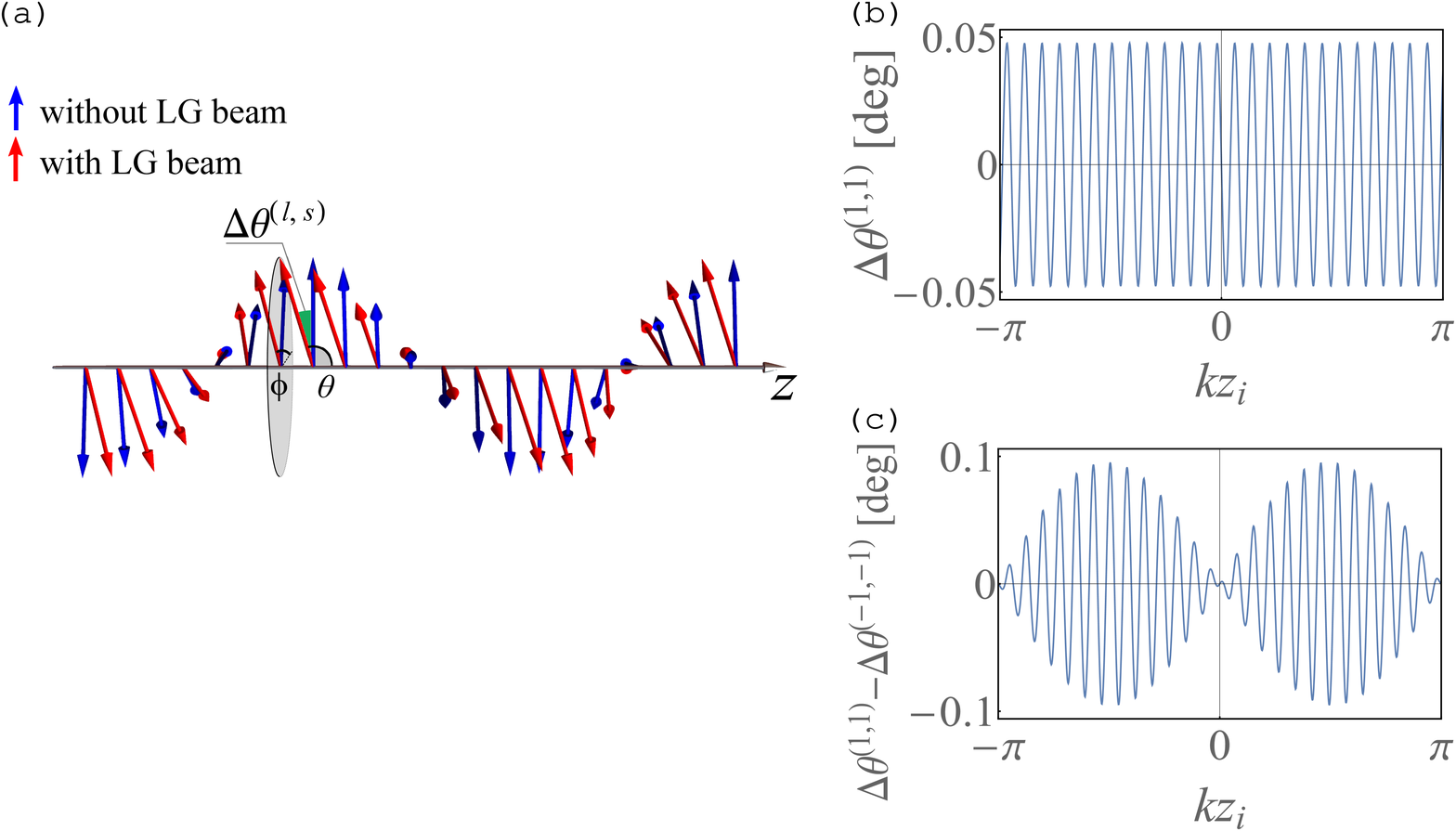}
\par\end{centering}
\caption{(a) Schematic view of the modulated magnetization induced by the LG beam. Each local spin is characterized by the polar angle $\theta$
and the azimuth angle $\phi$ with respect to the $z$-axis. If the optical vortex field is absent, the angle $\theta$ is uniformly $90^{\circ}$ (blue arrow). Under the optical vortex radiation, the spins are tilted
(red arrows). $\Delta\theta^{\left(l,s\right)}$ is the deviation of the polar angle from $90^{\circ}$ in the presence of the vortex field with $\left(l,s\right)$. (b) $\Delta\theta^{\left(1,1\right)}$ is plotted along the helical axis ($z$-axis). The parameters are $a=1.212$ nm, $D/J=0.16$, $\alpha=0.3$, $\lambda=1240$ nm, and $\mathcal{B}_{0}=10D/J$. The spatial pitch of the oscillation corresponds that of the helical magnetic structure $4\pi^2 k a J/D \sim 0.241\,\text{rad}$.
(c) The difference between $\Delta\theta^{\left(1,1\right)}$ and $\Delta\theta^{\left(-1,-1\right)}$ is plotted. }
\label{fig:theta}
\end{figure}

We show in figures \ref{fig:theta} the modulation of the steady-state magnetization along the helical axis ($z$-axis) under the optical vortex radiation with $\left(l,s\right)=\left(1,1\right)$. In the numerical simulation, we choose the following parameters: the lattice constant $a=1.212$ nm \citep{RN303}, $D/J=0.16$ \citep{RN364},
$\alpha=0.3$, $w_{0}=1240$ nm, the wavelength $\lambda=2\pi/k=1240$nm, and $\mathcal{B}_{0}=10D/J$. We write the local spin direction as $\hat{\bm{n}}_{i}=\left(1,\theta\left(z_{i}\right),\phi\left(z_{i}\right)\right)$, where $\theta$ is the polar angle and $\phi$ is the azimuth angle with respect to the $z$-axis (see figure \ref{fig:theta}(a)). When the optical vortex field is absent, the angle $\theta$ is uniformly $90^{\circ}$ with $\phi$ rotating around the $z$-axis. In figure \ref{fig:theta}(b), the deviation of the polar angle from $90^{\circ}$ ($\Delta\theta^{\left(l,s\right)}$),
corresponding to the light with $\left(l,s\right)=\left(1,1\right)$ is plotted. One can see that $\Delta\theta^{\left(1,1\right)}$ oscillates with the spiral pitch of the original chiral magnetic order. For the optical vortex radiation with $\left(l,s\right)=\left(-1,-1\right)$,
the deviation $\Delta\theta^{\left(-1,-1\right)}$ exhibits the oscillation with the same period as $\Delta\theta^{\left(1,1\right)}$. In figure
\ref{fig:theta}(c), we show the difference between $\Delta\theta^{\left(1,1\right)}$
and $\Delta\theta^{\left(-1,-1\right)}$. One can see that the envelop of the difference reflects the oscillation of $\bm{b}_{i}^{\left(1,1\right)}-\bm{b}_{i}^{\left(-1,-1\right)}$. These results clearly show that the coupling between the LG beam and the spins can modulate
the rigid chiral structure with the spatial structure of light. Such a spiral-pitch modulation of the macroscopic magnetic order in the helical axis is difficult to achieve by applying the external magnetic field. The results imply the possibility of the light with finite OAM to control the magnetic order.

\begin{figure}
\begin{centering}
\includegraphics[width=0.95\linewidth]{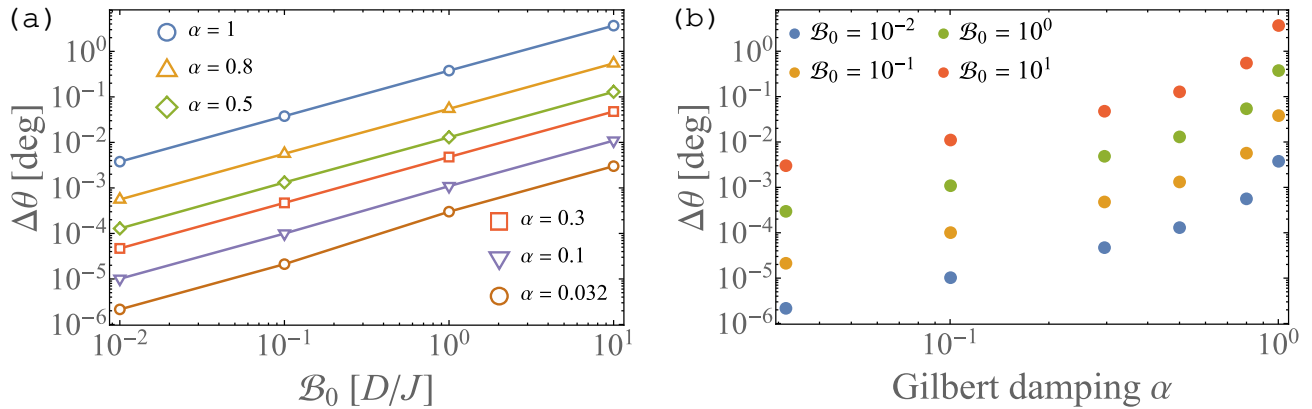}
\par\end{centering}
\caption{(a) Dependence of the polar angle deviation $\Delta\theta$ on the intensity of the vortex-spin coupling $\mathcal{B}_{0}=10^{m}D/J$ $(m=-2,-1,0,1)$. The other parameters are the same in figure \ref{fig:theta}(b). (b) Dependence of $\Delta\theta$ on the Gilbert damping constant
$\alpha=0.032,0.1,0.3,0.5,0.8$, and $1$ with the other parameters set to be the same in figure \ref{fig:theta}(b).}
\label{fig:Dependence}
\end{figure}

In figure \ref{fig:Dependence}(a), we present the peak deviation of the polar angle $\Delta\theta$ by changing only the spin-light coupling strength $\mathcal{B}_{0}=10^{m}D/J$ $(m=-2,-1,0,1)$, regardless of $\left(l,s\right)$. One can see that it is linearly proportional to the optical vortex-induced spin-spin interaction, i.e., proportional to the intensity of the LG beam. To our best knowledge, the precise value of the Gilbert damping constant of $\ce{CrNb3S6}$ is not reported. Thus, we investigated the dependence of the magnetic structure on this constant. In figure \ref{fig:Dependence}(b), the peak value of $\Delta\theta$ is shown by changing only the Gilbert
damping constant $\alpha=0.032,0.1,0.3,0.5,0.8$, and $1$. They show that the peak magnitude of the deviation monotonically increases with the Gilbert damping constant.

In our simulation, we have assumed that the helical magnetic structure and the optical vortex are coaxial. The optical selection rule by the optical vortex is still valid even though the target ion does not lie at the vortex center \citep{RN358,RN354,RN533}. Thus, our result is robust against the slight axis misalignment. In addition, when the light is obliquely incident with angle $\beta$, the vortex-induced coupling becomes $\bm{b}_{i}^{\left(\pm1,\pm1\right)}=-\cos \left[ k \left( z_{i}+a/2\right) \right] \hat{\bm{e}}^{x}\pm\sin \left[ k \left( z_{i}+a/2\right) \right]\cos\beta\hat{\bm{e}}^{y} \pm\sin \left[ k \left( z_{i}+a/2\right) \right]\sin\beta\hat{\bm{e}}^{z}$. Although the coupling is reduced depending on the angle of incidence, it is possible to confirm the change in the magnetic structure accordingly. Besides, the $z$-component of $\bm{b}_{i}^{\left(\pm1,\pm1\right)}$ changes the Heisenberg exchange interaction so little that it can be ignored.

In the present setup, the helimagnet lies at the center axis of the optical vortex. Because the electric field intensity becomes weak at the vortex center, the vortex-electron coupling is considerably small. However, in the ion trap system, the excitation reflecting the transition selection rule is confirmed even near the center of the vortex axis \citep{RN358}. We think that the results obtained here can be verified in this system by controlling the shape and intensity of the optical vortex beam. The coupling strength depends on the beam waist of the vortex beam. For the optical vortex with $l=s=\pm 1$, the optical transition matrix is approximately inversely proportional to the waist (see the Supplementary Material). Thus, in our calculation, we consider the vortex beam waist narrowed almost to the diffraction limit. On the other hand, in ion trap systems, optical transitions that reflect the OAM of light have been confirmed even when the ion position is shifted from the center of the vortex \citep{RN533}. For the derivation of the optical vortex-induced interaction, it is sufficient to have the optical selection rule modified by the OAM of light. Thus, even if the 1D helimagnet is arranged off the central axis of the vortex, our results will not be significantly different. As long as the material absorbs the light with OAM, the various magnetic structure mediated by the itinerant electrons can be controlled.

There is one more thing to note about the system parameters to verify our work. In general, a strongly-driven system with an ac field such as lasers may increase the system temperature via coupling to its environment. As is mentioned in section III, the vortex-induced interaction essentially requires the finite absorption line width $\delta$. Then, the interaction should be affected by this heating effect to some extent. Thus, in this work, we have considered the parameter regime where the optical vortex-induced coupling is sufficiently smaller than the DM interaction, and is treated sufficiently within the perturbation regime. Seeing equations (\ref{eq:vprime}) and (\ref{eq:B}), one may be concerned that the third-order perturbation looks unreasonably large due to the factor $\delta^{-1}$. However, in fact, even when irradiating the strongest optical vortex beam among the parameters assumed here ($\mathcal{B}_{0}=10D/J$), the magnitude of the coefficient of the third term on the right-hand side of equation (\ref{eq:EffectiveHamiltonian2}) is at most 0.01 that of the second term (DM interaction term). Here, the relaxation time is determined by the phenomenological line width $\delta$ for the optical absorption. To evaluate the heating effect in more detail, we need to microscopically take into account the coupling with a heat bath. There are many approaches that treat the dissipation effects in driving systems, e.g., Green's function method \cite{Aoki2014,Haug2008,KeldyshTsuji2008,KeldyshEmanuele2013} and quantum master equation method \cite{RN616,Breuer2000,DensityMatrixDiehl2008,DensityMatrixSchmidt2011,DensityMatrixCuetara2015,Ikeda2020}.

\section{Conclusion}
We have formulated the spin-spin interaction induced by twisted light, i.e., optical vortex, in a 1-D chiral helimagnet such as chromium-intercalated niobium disulfide $\ce{CrNb3S6}$. We have also shown by the LLG numerical calculation that such an optical vortex-induced interaction can modulate the chiral magnetic order to the unconventional phase structure.

Switchable control of magnetic orders is a fundamental and critical element for the evolution of spintronic devices. There are various attempts to control one- or two-dimensional chiral magnetic domain structure with cavity optomechanical system \citep{RN37}, electric currents \citep{RN548}, and electromagnetic fields \citep{RN612,RN613,RN614,RN615}. Here, we have microscopically derived a new type of spin-spin interaction induced by the optical vortex absorption. The vortex-induced interaction enables us to control the 1-D chiral spin order due to a completely different way. This interaction is attributed to the optical transition process between the $d$-orbitals, which requires the presence of the OAM of light. In addition, the spacial structure of the optical vortex beam provides a kind of chiral couplings with the magnetic structure. There are many degrees of freedom for optical vortex parameters, e.g., intensity, beam radius, polarization, and OAM, and these degrees of freedom can be helpful to such a control scheme. For instance, spatially-confined light can modulate magnetic structures locally. The local vortex-induced modulation is expected to affect the transport property of the quasiparticles, such as spin waves. The accessibility to the spin-wave dynamics in the chiral helimagnet has a possibility to open a route for new spintronic functionalities in device applications.

True completion of this study can present a principle for magneto-optics in a broader framework. Therefore, by clarifying the optical response of the magnetic order theoretically and suggesting the potential to control the magnetic order with light into experimental studies, we can blaze a trail in the field of opto-spintronics research.

\section*{Acknowledgement}
This was supported by JSPS KAKENHI Grant-in-Aid for Scientific Research (C) No. 19K03688, and No. JP16H06504 in Scientific Research on Innovative Areas ``Nano-Material Optical-Manipulation''. Y.G. was supported by the support by JSPS KAKENHI Grant Number JP19J15165, and Graduate Course for System-inspired Leaders in Material Science of JSPS. N.Y. was supported by a grant from Inamori Foundation, and Faculty Innovation (FI) fellowship support grant in Graduate School of Engineering, Osaka Prefecture University.

\bibliographystyle{iopart-num}
\bibliography{refs}

\end{document}


\global\long\def\theequation{S\arabic{equation}}%

\title{Supplementary Material: Twisted light-induced spin-spin interaction
in a chiral helimagnet}
\author{Yutaro Goto}
\email{goto-2@pe.osakafu-u.ac.jp}

\affiliation{Department of Physics and Electronics, Osaka Prefecture University,
Osaka 599-8531, Japan}
\author{Hajime Ishihara}
\affiliation{Department of Physics and Electronics, Osaka Prefecture University,
Osaka 599-8531, Japan}
\affiliation{Department of Materials Engineering Science, Osaka University, Osaka
560-8531, Japan}
\affiliation{Quantum Information and Quantum Biology Division, Institute for Open
and Transdisciplinary Research Initiatives, Osaka University, Osaka
560-8531, Japan}
\author{Nobuhiko Yokoshi}
\email{yokoshi@pe.osakafu-u.ac.jp}

\affiliation{Department of Physics and Electronics, Osaka Prefecture University,
Osaka 599-8531, Japan}

\maketitle
In this Supplementary Material, we show the derivation of the effective Hamiltonian (equation (13)), which is shown in section III of the main text. We also detail the optical transitions by the optical vortex shown in section IV.

\section{Effective spin-spin Hamiltonian}

We show the derivation of the effective
spin-spin Hamiltonian for the localized spin by the perturbative calculation, from tight-binding model Hamiltonian including the interaction among the $d$-electron, local moments, and optical vortex radiation.

\subsection{Tight-binding model Hamiltonian}

We employ a tight-binding model Hamiltonian in the whole system as for the electronic spin system in the chiral helimagnet and introduce the vector potential through the
on-site interband excitation. The Hamiltonian includes the $d$-electrons
energy, Hund's coupling, and hopping energy between adjacent sites
as follows;
\begin{align}
\mathcal{H}_{\text{ch}} & =\mathcal{H}_{0}+\mathcal{H}_{\text{hop}}+\mathcal{H}_{\text{SO}},\label{eq:SystemHamiltionian}
\end{align}
where we define
\begin{align}
\mathcal{H}_{0} & =\sum_{i,\mu,\gamma}E_{\mu}c_{i\mu\gamma}^{\dagger}c_{i\mu\gamma}-J_{\text{H}}\sum_{i,\mu,\gamma,\gamma^{\prime}}\bm{S}_{i}\cdot\bm{s}_{i},\\
\mathcal{H}_{\text{hop}} & =-\Gamma\sum_{i,j,\mu,\gamma,\gamma^{\prime}}c_{i\mu\gamma}^{\dagger}c_{j\mu\gamma^{\prime}},\label{eq:Hamiltonian_hopping}\\
\mathcal{H}_{\text{SO}} & =\lambda\sum_{i}\bm{L}_{i}\cdot\bm{s}_{i},
\end{align}
with the energy of $d$-electrons $E_{\mu}$, the annihilation operator
$c_{i\mu\gamma}$ such that the field operator of the electron is
$\psi_{\mu\gamma}\left(\bm{r}\right)=\Sigma_{i}\phi_{i\mu}\left(\bm{r}\right)c_{i\mu\gamma}$
where $\phi_{i\mu}\left(\bm{r}\right)$ are the Wannier function centered
on the site $\bm{R}_{i}$. The indices $\left\{ \mu,\nu\right\} $
and $\left\{ \gamma,\gamma^{\prime}\right\} $ indicate the orbit
and spin. Here we consider $a_{1}$- and $e$-symmetry band for
the orbit. The constants $J_{\text{H}},\Gamma$ and $\lambda$ denote the Hund\textquoteright s coupling energy between the electron spin $\bm{s}_{i}$ and the localized spin $\bm{S}_{i}$ with $S=\abs{\bm{S}_{i}}=3/2$, the hopping energy, and the spin-orbit coupling, respectively. We have neglected the spin-orbit interaction between the local momenta $\bm{L}_{i}\cdot\bm{S}_{i}$. It contributes to the interband transition with spin-flipping whose transition energy is much larger than that by $\bm{L}_{i}\cdot\bm{s}_{i}$ within the Hund's rule. In addition,
considering the effect of the optical vortex absorption, the $\bm{L}_{i}\cdot\bm{S}_{i}$ term should be higher-order process.

In the absence of the spin-orbit interaction $\mathcal{H}_{\text{SO}}$
and assuming the condition $\Gamma \ll J_{\text{H}}$, the hopping
$\mathcal{H}_{\text{hop}}$ can be represented by employing the projection method as \citep{RN174}
\begin{align}
\mathcal{H}_{\text{hop}} & =-\Gamma\sum_{i,j,\mu,\gamma,\gamma^{\prime}}c_{i\mu\gamma}^{\dagger}\bra{\gamma}P_{i}^{0}P_{j}^{0}\ket{\gamma^{\prime}}c_{j\mu\gamma^{\prime}},\label{eq:Hamiltonian_hopping2}
\end{align}
where $P_{i}^{0}$ is the projection operator at the $i$ site;
\begin{align}
P_{i}^{0} & =\frac{\bm{S}_{i}\cdot\bm{\sigma}+\left(S+1\right)\sigma_{0}}{2S+1}.\label{eq:ProjectionOperator0}
\end{align}
Here $\mu$ denotes the orbital state with the lowest possible energy with the Hund\textquoteright s coupling also considered. The Pauli matrix denotes the spin of itinerant electrons.

To introduce the interaction with the optical vortex, we introduce the `vector potential' in the presence of the spin-orbit interaction \citep{RN29};
\begin{align*}
\bm{\chi}^{\left(l,s\right)}\left(\bm{r},t\right)=-e\bm{A}^{\left(l,s\right)}\left(\bm{r},t\right)-\left(\mu_{\text{B}}/2c^{2}\right)\bm{\sigma}\times\bm{E}^{\left(l,s\right)}\left(\bm{r},t\right),
\end{align*}
with ${\bm{\sigma}}$ being the Pauli matrix. Here,
$\mu_{\text{B}}$ is the Bohr magneton, and $\bm{E}^{\left(l,s\right)}\equiv i\omega\bm{A}^{\left(l,s\right)}-\nabla\varphi^{\left(l,s\right)}$.
We additionally introduce the vector potential through the on-site interband excitation by the optical vortex radiation,
i.e., $\mathcal{H}^{\prime}\left(t\right)=\mathcal{H}_{\text{ch}}+\mathcal{H}^{\bm{\chi}}\left(\bm{r},t\right)$, where
\begin{align*}
\mathcal{H}^{\bm{\chi}}\left(\bm{r},t\right) & =\frac{1}{2m}\sum_{i,\mu\neq\nu,\gamma,\gamma^{\prime}}\mel*{i\mu\gamma}{\left(\bm{p}\cdot\bm{\chi}^{\left(l,s\right)}\left(\bm{r},t\right)+\bm{\chi}^{\left(l,s\right)}\left(\bm{r},t\right)\cdot\bm{p}+\text{H.c.}\right)}{i\nu\gamma^{\prime}}c_{i\mu\gamma}^{\dagger}c_{i\nu\gamma^{\prime}}\\
 & =\frac{i}{\hbar}\sum_{i,\mu\neq\nu,\gamma,\gamma^{\prime}}E_{\mu\nu}\mel*{i\mu\gamma}{\left(\bm{r}\cdot\bm{\chi}^{\left(l,s\right)}\left(\bm{r},t\right)+\text{H.c.}\right)}{i\nu\gamma^{\prime}}c_{i\mu\gamma}^{\dagger}c_{i\nu\gamma^{\prime}},
\end{align*}
with $m,\bm{p}$ and $E_{\mu\nu}$ being the mass, momentum of electron,
and transition energy. In addition, we have defined the matrix element
for some operator $\bm{\mathcal{O}}$ as $\mel{i\mu\gamma}{\bm{\mathcal{O}}}{j\nu\gamma^{\prime}}=\mel{\gamma}{\int d\bm{r}\phi_{i\mu}^{*}\left(\bm{r}\right)\bm{\mathcal{O}}\left(\bm{r}\right)\phi_{j\nu}\left(\bm{r}\right)}{\gamma^{\prime}}$.
The electric field is assumed to be resonant with $E_{g}$, where
$E_{g}=E_{e}-E_{a_{1}}+SJ_{\text{H}}$ is the gap energy including
the Hund's coupling.

\subsection{Treatment of the perturbative Hamiltonian}

We derive the effective model considering the perturbations with respect
to the hopping $\mathcal{H}_{\text{hop}}$, spin-orbit interaction
$\mathcal{H}_{\text{SO}}$, and optical vortex absorption $\mathcal{H}^{\bm{\chi}}\left(t\right)$.
The unperturbed Hamiltonian is $\mathcal{H}_{0}$, and the itinerant
electrons in the lower band ($a_{1}$) have their spins parallel to
the localized spins in the unperturbed ground state. Thus, we impose the constraint not to break the Hund rule throughout the perturbation calculation. Here we consider the time evolution operator in the interaction picture
\begin{align}
U\left(t,0\right) & =\mathcal{T}\exp\left\{ -\frac{i}{\hbar}\int_{0}^{t}\left(\mathcal{H}_{\text{hop}}+\mathcal{H}_{\text{SO}}+\mathcal{H}^{\bm{\chi}}\left(t^{\prime}\right)\right)dt^{\prime}\right\} ,
\end{align}
where $\mathcal{T}$ is the time-ordering operator. The zero-th order
$U_{\text{I}}^{\left(0\right)}\left(t;0\right)$ is identity. We consider
only the energy terms $\mathcal{E}^{\left(\zeta\right)}$ from one
to three in $\zeta$ are treated as the $\zeta$-th-order perturbation.

\paragraph*{First-order perturbation}

The first-order term is just $\mathcal{H}_{\text{hop}}$, then

\begin{align*}
\mathcal{H}^{\left(1\right)} & =-\Gamma\sum_{i,j,\mu,\gamma,\gamma^{\prime}}c_{i\mu\gamma}^{\dagger}\bra{\gamma}P_{i}^{0}P_{j}^{0}\ket{\gamma^{\prime}}c_{j\mu\gamma^{\prime}}.
\end{align*}
The term indicates that the localized spins are aligned for itinerant
electrons to have more extended wave forms. Here we assume $\bm{S}_{j}$
being a classical vector $S\left(\cos\varphi\sin\vartheta,\sin\varphi\sin\vartheta,\cos\vartheta\right)$,
then
\begin{align}
P_{j}^{0} & =\frac{1}{2S+1}\begin{pmatrix}S\cos\vartheta_{j}+S+1 & S\sin\vartheta_{j}e^{-i\varphi_{j}}\\
S\sin\vartheta_{j}e^{i\varphi_{j}} & -S\cos\vartheta_{j}+S+1
\end{pmatrix}.\label{eq:ProjectionOperator_angle}
\end{align}
Diagonalizing $P_{j}$, we obtain the eigensystem,

\begin{align*}
\tilde{P}_{j} & =M_{j}^{-1}P_{j}^{0}M_{j}=\begin{pmatrix}1 & 0\\
0 & 1/\left(2S+1\right)
\end{pmatrix},
\end{align*}
and
\[
M_{j}=\begin{pmatrix}e^{-i\frac{\varphi_{j}}{2}}\cos\frac{\vartheta_{j}}{2} & e^{-i\frac{\varphi_{j}}{2}}\sin\frac{\vartheta_{j}}{2}\\
e^{i\frac{\varphi_{j}}{2}}\sin\frac{\vartheta_{j}}{2} & -e^{i\frac{\varphi_{j}}{2}}\cos\frac{\vartheta_{j}}{2}
\end{pmatrix}.
\]
Thus, the Hamiltonian can be written by those matrices
\begin{align*}
\mathcal{H}^{\left(1\right)} & =-\Gamma\sum_{i,j,\mu,\gamma,\gamma^{\prime}}c_{i\mu\gamma}^{\dagger}\bra{\gamma}M_{i}\tilde{P}_{j}M_{i}^{-1}M_{j}\tilde{P}_{j}M_{j}^{-1}\ket{\gamma^{\prime}}c_{j\mu\gamma^{\prime}}.
\end{align*}
Here we consider the low energy state only, and neglect the higher inhomogeneous hopping energy which may cause charge accumulation. The Hamiltonian reduces to
\begin{align}
\mathcal{H}^{\left(1\right)} & \approx-t\sum_{i,j,\mu}\left[\cos\varphi_{ij}\cos\left(\frac{\vartheta_{ij}}{2}\right)\right]d_{i\mu-}^{\dagger}d_{j\mu-}+\text{H.c.},\label{eq:FerromagneticHamiltonian}
\end{align}
where $\varphi_{ij}=\varphi_{i}-\varphi_{j},\vartheta_{ij}=\vartheta_{i}-\vartheta_{j}$
and
\begin{align*}
d_{j\mu-} & =e^{i\frac{\varphi_{j}}{2}}\cos\frac{\vartheta_{j}}{2}c_{j\mu\uparrow}+e^{-i\frac{\varphi_{j}}{2}}\sin\frac{\vartheta_{j}}{2}c_{j\mu\downarrow},
\end{align*}
is the field operator of the electrons in the eigenstate of lower
energy. Since the corrected hopping energy is proportional to $\cos\left(\vartheta_{ij}/2\right)$
for the lower energy, we can reach the equation \eqref{eq:FerromagneticHamiltonian},
which favors the ferromagnetic state.

\paragraph*{Second-order perturbation}

For the second-order term, it depends on the transition matrix of two products from each of the hopping $\mathcal{H}_{\text{hop}}$, spin-orbit interaction $\mathcal{H}_{\text{SO}}$, and optical vortex absorption
$\mathcal{H}^{\bm{\chi}}\left(t\right)$. The only ones for which transitions that exist are the following;

\begin{align*}
 & \mel{i,\mu,\gamma}{U_{\text{I}}^{\left(2\right)}\left(t;0\right)}{j,\mu,\gamma^{\prime}}\\
= & -\frac{1}{\hbar^{2}}\int_{_{0}}^{t}dt_{1}\int_{0}^{t_{1}}dt_{2}\left(\mel{i,\mu,\gamma}{\mathcal{H}_{\text{hop, I}}^{\prime}\left(t_{1}\right)\mathcal{H}_{\text{SO, I}}\left(t_{2}\right)}{j,\mu,\gamma^{\prime}}+\mel{i,\mu,\gamma}{\mathcal{H}_{\text{SO, I}}\left(t_{1}\right)\mathcal{H}_{\text{hop, I}}^{\prime}\left(t_{2}\right)}{j,\mu,\gamma^{\prime}}\right)\\
 & -\frac{1}{\hbar^{2}}\int_{_{0}}^{t}dt_{1}\int_{0}^{t_{1}}dt_{2}\left(\mel{i,\mu,\gamma}{\mathcal{H}_{\text{hop, I}}^{\prime}\left(t_{1}\right)\mathcal{H}_{\text{I}}^{\bm{\chi}}\left(t_{2}\right)}{j,\mu,\gamma^{\prime}}+\mel{i,\mu,\gamma}{\mathcal{H}_{\text{I}}^{\bm{\chi}}\left(t_{1}\right)\mathcal{H}_{\text{hop, I}}^{\prime}\left(t_{2}\right)}{j,\mu,\gamma^{\prime}}\right)\\
= & \frac{i}{\hbar}\left[\frac{E_{g}}{E_{g}-\hbar\omega-i\delta}\left(\frac{e^{i\left(\hbar\omega+i\delta\right)t/\hbar}-1}{\hbar\omega+i\delta}-\frac{e^{iE_{g}t/\hbar}-1}{E_{g}}\right)+\frac{E_{g}}{E_{g}+\hbar\omega-i\delta}\left(\frac{e^{i\left(-\hbar\omega+i\delta\right)t/\hbar}-1}{-\hbar\omega+i\delta}-\frac{e^{iE_{g}t/\hbar}-1}{E_{g}}\right)\right]\\
 & \times\left(1-\delta_{ij}\right)\left(1-\delta_{\mu\nu}\right)\sum_{\gamma^{\prime\prime}}H_{\text{hop}}^{\prime}\left(i\mu\gamma;j\nu\gamma^{\prime\prime}\right)H_{\text{dip}}^{\bm{\chi}}\left(j\nu\gamma^{\prime\prime};j\mu\gamma^{\prime}\right)\\
 & -\frac{i}{\hbar}\left[\left(\frac{e^{i\left(\hbar\omega+i\delta\right)t/\hbar}-1}{\hbar\omega+i\delta}-\frac{e^{i\left(E_{g}+\hbar\omega+i\delta\right)t/\hbar}-1}{E_{g}+\hbar\omega+i\delta}\right)+\left(\frac{e^{i\left(-\hbar\omega+i\delta\right)t/\hbar}-1}{-\hbar\omega+i\delta}-\frac{e^{i\left(E_{g}-\hbar\omega+i\delta\right)t/\hbar}-1}{E_{g}-\hbar\omega+i\delta}\right)\right]\\
 & \times\left(1-\delta_{ij}\right)\left(1-\delta_{\mu\nu}\right)\sum_{\gamma^{\prime\prime}}H_{\text{dip}}^{\bm{\chi}}\left(i\mu\gamma;i\nu\gamma^{\prime\prime}\right)H_{\text{hop}}^{\prime}\left(i\nu\gamma^{\prime\prime};j\mu\gamma^{\prime}\right)\\
 & -\frac{1}{E_{g}}\left(\frac{i}{\hbar}t-\frac{e^{iE_{g}t/\hbar}-1}{E_{g}}\right)\sum_{\gamma^{\prime\prime}}\left(H_{\text{hop}}^{\prime}\left(i\mu\gamma;j\nu\gamma^{\prime\prime}\right)H_{\text{SO}}\left(j\nu\gamma^{\prime\prime};j\mu\gamma^{\prime}\right)+H_{\text{SO}}\left(i\mu\gamma;i\nu\gamma^{\prime\prime}\right)H_{\text{hop}}^{\prime}\left(i\nu\gamma^{\prime\prime};j\mu\gamma^{\prime}\right)\right),
\end{align*}
where
\begin{align*}
\mathcal{H}_{\text{hop}}^{\prime} & =-\Gamma\sum_{i,j,\mu,\nu,\gamma,\gamma^{\prime}}\left(1-\delta_{ij}\right)c_{i\mu\gamma}^{\dagger}\bra{\gamma}P_{i}^{0}P_{j}^{0}\ket{\gamma^{\prime}}c_{j\nu\gamma^{\prime}},
\end{align*}
and the transition amplitude
\begin{align*}
H_{\text{hop}}^{\prime}\left(i\mu\gamma;j\nu\gamma^{\prime}\right) & =\mel{i,\mu,\gamma}{\mathcal{H}_{\text{hop}}^{\prime}}{j,\nu,\gamma^{\prime}},\\
H_{\text{SO}}\left(i\mu\gamma;i\nu\gamma^{\prime}\right) & =\mel*{i\mu\gamma}{\mathcal{H}_{\text{SO}}}{i\nu\gamma^{\prime}},\\
H_{\text{dip}}^{\bm{\chi}}\left(i\mu\gamma;i\nu\gamma^{\prime}\right) & =-\frac{i\hbar}{E_{g}}\mel*{i\mu\gamma}{\mathcal{H}^{\bm{\chi}}\left(\bm{r}\right)}{i\nu\gamma^{\prime}},
\end{align*}
with $\delta$ being the phenomenological line width for the optical
transition. Here $\mathcal{H}^{\bm{\chi}}\left(\bm{r}\right)$ is
$\mathcal{H}^{\bm{\chi}}\left(\bm{r},t\right)$ with the temporal
dependence omitted. We relax the restriction of the band transition
we put on $\mathcal{H}_{\text{hop}}$ in equation \eqref{eq:Hamiltonian_hopping}
because other orbitals can be accessed via the spin-orbit interaction.
Assuming that the perturbation energy is much smaller than the bandgap
energy, we use secular approximation and ignore the temporal oscillation
term like the Rabi oscillation under the resonant condition \citep{RN616,RN617}. In addition,
the period of the neglected oscillation $\sim2\pi\hbar/E_{g}$ is
much shorter than the time step of the subsequent simulation. Such
a oscillation should be averaged out during the time step. It thus
becomes
\begin{align*}
 & \mel{i,\mu,\gamma}{U_{\text{I}}^{\left(2\right)}\left(t;0\right)}{j,\mu,\gamma^{\prime}}\\
\approx & -\frac{i}{\hbar E_{g}}t\left(1-\delta_{ij}\right)\left(1-\delta_{\mu\nu}\right)\sum_{\gamma^{\prime\prime}}\left(H_{\text{hop}}^{\prime}\left(i\mu\gamma;j\nu\gamma^{\prime\prime}\right)H_{\text{SO}}\left(j\nu\gamma^{\prime\prime};j\mu\gamma^{\prime}\right)+H_{\text{SO}}\left(i\mu\gamma;i\nu\gamma^{\prime\prime}\right)H_{\text{hop}}^{\prime}\left(i\nu\gamma^{\prime\prime};j\mu\gamma^{\prime}\right)\right).
\end{align*}
Since $\mel*{\mu}{\bm{L}_{i}}{\nu}\ \left(\mu\neq\nu\right)$ is purely imaginary, we obtain the second-order perturbation energy as
\begin{align*}
\mathcal{E}^{\left(2\right)} & =\sum_{i,j,\mu,\nu,\gamma,\gamma^{\prime},\gamma^{\prime\prime}}\frac{1}{E_{g}}\left(1-\delta_{ij}\right)\left(1-\delta_{\mu\nu}\right)\left[\begin{aligned}H_{\text{hop}}^{\prime}\left(i\mu\gamma;j\nu\gamma^{\prime\prime}\right)H_{\text{SO}}\left(j\nu\gamma^{\prime\prime};j\mu\gamma^{\prime}\right)+H_{\text{SO}}\left(i\mu\gamma;i\nu\gamma^{\prime\prime}\right)H_{\text{hop}}^{\prime}\left(i\nu\gamma^{\prime\prime};j\mu\gamma^{\prime}\right)\\
H_{\text{hop}}^{\prime}\left(i\mu\gamma;j\nu\gamma^{\prime\prime}\right)H_{\text{SO}}\left(i\nu\gamma^{\prime\prime};i\mu\gamma^{\prime}\right)+H_{\text{SO}}\left(j\mu\gamma;j\nu\gamma^{\prime\prime}\right)H_{\text{hop}}^{\prime}\left(i\nu\gamma^{\prime\prime};j\mu\gamma^{\prime}\right)
\end{aligned}
\right].
\end{align*}
We substitute
$H_{\text{hop}}^{\prime}\left(i\mu\gamma;j\nu\gamma^{\prime}\right)$
and $H_{\text{SO}}\left(i\mu\gamma;i\nu\gamma^{\prime}\right)$, and
simplify the terms by using the commutator. Thus, it can be rewritten
as the perturbation Hamiltonian in the second order \citep{RN416};
\begin{align}
  \mathcal{H}^{\left(2\right)} & =\frac{\Gamma\lambda}{E_{g}}\sum_{i,j,\mu,\nu,\gamma,\gamma^{\prime}}\left(1-\delta_{ij}\right)\left(1-\delta_{\mu\nu}\right)c_{i\mu\gamma}^{\dagger}\bra{\gamma}\left(\mel*{a_{1}}{\bm{L}_{i}}{e}\cdot P_{i}^{0}\left[\bm{\sigma}_{j},P_{j}^{0}\right]+\mel*{a_{1}}{\bm{L}_{i}}{e}\cdot\left[\bm{\sigma}_{i},P_{i}^{0}\right]P_{j}^{0}\right)\ket{\gamma^{\prime}}c_{j\nu\gamma^{\prime}}\nonumber \\
   & =-\frac{\Gamma}{2S+1}\sum_{i,j,\mu,\nu,\gamma,\gamma^{\prime}}\left(1-\delta_{ij}\right)\left(1-\delta_{\mu\nu}\right)c_{i\mu\gamma}^{\dagger}\bra{\gamma}\left(P_{i}^{0}\bm{\sigma}_{j}\cdot\bm{v}_{j}+\bm{\sigma}_{i}\cdot\bm{v}_{i}P_{j}^{0}\right)\ket{\gamma}c_{j\nu\gamma^{\prime}},\label{eq:2nd-orderHamiltonian}
  \end{align}
where
\begin{align}
\bm{v}_{i} & =-\frac{i\lambda}{E_{g}}\mel{a_{1}}{\bm{L}_{i}}{e}\times\bm{S}_{i}.\label{eq:2nd_v_term}
\end{align}
The vector \eqref{eq:2nd_v_term} corresponds to equation (11) in the main text.

\paragraph*{Third-order perturbation}

For the third-order term, the transitions that exist are for the following;
\begin{align*}
 & \mel{i,\mu,\gamma}{U_{\text{I}}^{\left(3\right)}\left(t;0\right)}{j,\mu,\gamma^{\prime}}\\
= & +\frac{i}{\hbar^{3}}\int_{_{0}}^{t}dt_{1}\int_{0}^{t_{1}}dt_{2}\int_{0}^{t_{2}}dt_{3}\mel{i,\mu,\gamma}{\mathcal{H}_{\text{hop, I}}\left(t_{1}\right)\mathcal{H}_{\text{SO, I}}\left(t_{2}\right)\mathcal{H}_{\text{I}}^{\bm{\chi}}\left(t_{3}\right)}{j,\mu,\gamma^{\prime}}\\
 & +\frac{i}{\hbar^{3}}\int_{_{0}}^{t}dt_{1}\int_{0}^{t_{1}}dt_{2}\int_{0}^{t_{2}}dt_{3}\mel{i,\mu,\gamma}{\mathcal{H}_{\text{hop, I}}\left(t_{1}\right)\mathcal{H}_{\text{I}}^{\bm{\chi}}\left(t_{2}\right)\mathcal{H}_{\text{SO, I}}\left(t_{3}\right)}{j,\mu,\gamma^{\prime}}\\
 & +\frac{i}{\hbar^{3}}\int_{_{0}}^{t}dt_{1}\int_{0}^{t_{1}}dt_{2}\int_{0}^{t_{2}}dt_{3}\mel{i,\mu,\gamma}{\mathcal{H}_{\text{SO, I}}\left(t_{1}\right)\mathcal{H}_{\text{hop, I}}\left(t_{2}\right)\mathcal{H}_{\text{I}}^{\bm{\chi}}\left(t_{3}\right)}{j,\mu,\gamma^{\prime}}\\
 & +\frac{i}{\hbar^{3}}\int_{_{0}}^{t}dt_{1}\int_{0}^{t_{1}}dt_{2}\int_{0}^{t_{2}}dt_{3}\mel{i,\mu,\gamma}{\mathcal{H}_{\text{SO, I}}\left(t_{1}\right)\mathcal{H}_{\text{I}}^{\bm{\chi}}\left(t_{2}\right)\mathcal{H}_{\text{hop, I}}\left(t_{3}\right)}{j,\mu,\gamma^{\prime}}\\
 & +\frac{i}{\hbar^{3}}\int_{_{0}}^{t}dt_{1}\int_{0}^{t_{1}}dt_{2}\int_{0}^{t_{2}}dt_{3}\mel{i,\mu,\gamma}{\mathcal{H}_{\text{I}}^{\bm{\chi}}\left(t_{1}\right)\mathcal{H}_{\text{hop, I}}\left(t_{2}\right)\mathcal{H}_{\text{SO, I}}\left(t_{3}\right)}{j,\mu,\gamma^{\prime}}\\
 & +\frac{i}{\hbar^{3}}\int_{_{0}}^{t}dt_{1}\int_{0}^{t_{1}}dt_{2}\int_{0}^{t_{2}}dt_{3}\mel{i,\mu,\gamma}{\mathcal{H}_{\text{I}}^{\bm{\chi}}\left(t_{1}\right)\mathcal{H}_{\text{SO, I}}\left(t_{2}\right)\mathcal{H}_{\text{hop, I}}\left(t_{3}\right)}{j,\mu,\gamma^{\prime}}.
\end{align*}
We use the secular approximation under the resonant condition as the same way of the second-order perturbation. In addition, we consider the steady state where $e^{-\delta t}\sim0$ with the line width $\delta$. For instance, one can find
\begin{align*}
 & i\int_{_{0}}^{t}dt_{1}\int_{0}^{t_{1}}dt_{2}\int_{0}^{t_{2}}dt_{3}\mel{i,\mu,\gamma}{\mathcal{H}_{\text{hop, I}}\left(t_{1}\right)\mathcal{H}_{\text{I}}^{\bm{\chi}}\left(t_{2}\right)\mathcal{H}_{\text{SO, I}}\left(t_{3}\right)}{j,\mu,\gamma^{\prime}}\\
 & =\frac{i}{\hbar}\left[\frac{e^{i\left(\hbar\omega+i\delta\right)t/\hbar}-1}{\left(\hbar\omega+i\delta\right)^{2}}-\frac{it/\hbar}{\left(\hbar\omega+i\delta\right)}-\frac{e^{i\left(E_{g}+\hbar\omega+i\delta\right)t/\hbar}-1}{\left(E_{g}+\hbar\omega+i\delta\right)^{2}}+\frac{it/\hbar}{E_{g}+\hbar\omega+i\delta}\right.\\
 & \hphantom{=i}\ \ \left.+\frac{e^{i\left(-\hbar\omega+i\delta\right)t/\hbar}-1}{\left(-\hbar\omega+i\delta\right)^{2}}-\frac{it/\hbar}{\left(-\hbar\omega+i\delta\right)}-\frac{e^{i\left(E_{g}-\hbar\omega+i\delta\right)t/\hbar}-1}{\left(E_{g}-\hbar\omega+i\delta\right)^{2}}+\frac{it/\hbar}{E_{g}-\hbar\omega+i\delta}\right]\\
 & \hphantom{=}\times\left(1-\delta_{ij}\right)\left(1-\delta_{\mu\nu}\right)\sum_{\gamma^{\prime\prime},\gamma^{\prime\prime\prime}}H_{\text{hop}}\left(i\mu\gamma;j\mu\gamma^{\prime\prime}\right)H_{\text{dip}}^{\bm{\chi}}\left(j\mu\gamma^{\prime\prime};j\nu\gamma^{\prime\prime\prime}\right)H_{\text{SO}}\left(j\nu\gamma^{\prime\prime\prime};j\mu\gamma^{\prime}\right)\\
 & \approx\frac{it}{\hbar^{2}\delta}\left(1-\delta_{ij}\right)\left(1-\delta_{\mu\nu}\right)\sum_{\gamma^{\prime\prime},\gamma^{\prime\prime\prime}}H_{\text{hop}}\left(i\mu\gamma;j\mu\gamma^{\prime\prime}\right)H_{\text{dip}}^{\bm{\chi}}\left(j\mu\gamma^{\prime\prime};j\nu\gamma^{\prime\prime\prime}\right)H_{\text{SO}}\left(j\nu\gamma^{\prime\prime\prime};j\mu\gamma^{\prime}\right).
\end{align*}
The other terms reduce to zero or time-independent constant. Thus, we obtain the third-order perturbation energy
\begin{align*}
\mathcal{E}^{\left(3\right)}= & -\frac{1}{\hbar\delta}\sum_{i,j,\mu,\nu,\gamma,\gamma^{\prime}}\left(1-\delta_{ij}\right)\left(1-\delta_{\mu\nu}\right)\sum_{\gamma^{\prime\prime},\gamma^{\prime\prime\prime}}\left[\begin{aligned} & H_{\text{hop}}\left(j\mu\gamma;i\mu\gamma^{\prime\prime}\right)H_{\text{dip}}^{\bm{\chi}}\left(i\mu\gamma^{\prime\prime};i\nu\gamma^{\prime\prime\prime}\right)H_{\text{SO}}\left(i\nu\gamma^{\prime\prime\prime};i\mu\gamma^{\prime}\right)\\
+ & H_{\text{hop}}\left(i\mu\gamma;j\mu\gamma^{\prime\prime}\right)H_{\text{dip}}^{\bm{\chi}}\left(j\mu\gamma^{\prime\prime};j\nu\gamma^{\prime\prime\prime}\right)H_{\text{SO}}\left(j\nu\gamma^{\prime\prime\prime};j\mu\gamma^{\prime}\right)\\
+ & H_{\text{dip}}^{\bm{\chi}}\left(i\mu\gamma;i\nu\gamma^{\prime\prime}\right)H_{\text{hop}}\left(i\nu\gamma^{\prime\prime};j\nu\gamma^{\prime\prime\prime}\right)H_{\text{SO}}\left(i\nu\gamma^{\prime\prime\prime};i\mu\gamma^{\prime}\right)\\
+ & H_{\text{dip}}^{\bm{\chi}}\left(j\mu\gamma;j\nu\gamma^{\prime\prime}\right)H_{\text{hop}}\left(i\nu\gamma^{\prime\prime};j\nu\gamma^{\prime\prime\prime}\right)H_{\text{SO}}\left(j\nu\gamma^{\prime\prime\prime};j\mu\gamma^{\prime}\right)
\end{aligned}
\right].
\end{align*}
To symmetrize Hamiltonian for the site indices, we take the additional terms which are zero. It can be rewritten as the third-order perturbation Hamiltonian, substituting $H_{\text{hop}}\left(i\mu\gamma;j\mu\gamma^{\prime}\right),H_{\text{SO}}\left(i\mu\gamma;i\nu\gamma^{\prime}\right)$
and $H_{\text{dip}}^{\bm{\chi}}\left(i\mu\gamma;i\nu\gamma^{\prime}\right)$;
\begin{align}
  \mathcal{H}^{\left(3\right)} & =\frac{\Gamma\lambda}{2\hbar\delta}\sum_{i,j,\mu,\nu,\gamma,\gamma^{\prime}}c_{i\mu\gamma}^{\dagger}\left(1-\delta_{ij}\right)\left(1-\delta_{\mu\nu}\right)\nonumber \\
   & \quad\times\bra{\gamma}\left[\begin{aligned} & \mel*{a_{1}}{\bm{L}_{i}}{e}\cdot\left[\mel*{a_{1}}{\bm{r}_{i}\cdot\bm{\chi}_{i}^{\left(l,s\right)}}{e},P_{i}^{0}\right]\bm{\sigma}_{i}P_{j}^{0}\\
  + & \mel*{e}{\bm{L}_{i}}{a_{1}}\cdot\left[\mel*{e}{\bm{r}_{i}\cdot\bm{\chi}_{i}^{\left(l,s\right)}}{a_{1}},P_{i}^{0}\right]\bm{\sigma}_{i}P_{j}^{0}\\
  + & \mel*{a_{1}}{\bm{L}_{j}}{e}\cdot P_{i}^{0}\left[\mel*{a_{1}}{\bm{r}_{j}\cdot\bm{\chi}_{j}^{\left(l,s\right)}}{e},P_{j}^{0}\right]\bm{\sigma}_{j}\\
  + & \mel*{e}{\bm{L}_{j}}{a_{1}}\cdot P_{i}^{0}\left[\mel*{e}{\bm{r}_{j}\cdot\bm{\chi}_{j}^{\left(l,s\right)}}{a_{1}},P_{j}^{0}\right]\bm{\sigma}_{j}
  \end{aligned}
  \right]\ket{\gamma^{\prime}}c_{j\nu\gamma^{\prime}}\nonumber \\
   & =-\frac{\Gamma\lambda}{2\hbar\delta}\frac{4}{2S+1}\sum_{i,j,\mu,\nu,\gamma,\gamma^{\prime}}c_{i\mu\gamma}^{\dagger}\left(1-\delta_{ij}\right)\left(1-\delta_{\mu\nu}\right)\nonumber \\
   & \quad\times\bra{\gamma}\left[\begin{aligned}+ & P_{i}^{0}\left(\Im\mel*{a_{1}}{\bm{\mathcal{B}}_{j}^{\left(l,s\right)}}{e}\cdot\left(\mel*{a_{1}}{\bm{L}_{j}}{e}\times\bm{S}_{j}\right)\right)\sigma_{0}\\
  + & \left(\Im\mel*{a_{1}}{\bm{\mathcal{B}}_{i}^{\left(l,s\right)}}{e}\cdot\left(\mel*{a_{1}}{\bm{L}_{i}}{e}\times\bm{S}_{i}\right)\right)\sigma_{0}P_{j}^{0}\\
  + & iP_{i}^{0}\bm{\sigma}_{j}\cdot\left(\mel*{a_{1}}{\bm{L}_{j}}{e}\times\left(\Im\mel*{a_{1}}{\bm{\mathcal{B}}_{j}^{\left(l,s\right)}}{e}\times\bm{S}_{j}\right)\right)\\
  + & i\bm{\sigma}_{i}\cdot\left(\mel*{a_{1}}{\bm{L}_{i}}{e}\times\left(\Im\mel*{a_{1}}{\bm{\mathcal{B}}_{i}^{\left(l,s\right)}}{e}\times\bm{S}_{i}\right)\right)P_{j}^{0}
  \end{aligned}
  \right]\ket{\gamma^{\prime}}c_{j\nu\gamma^{\prime}}\nonumber \\
   & =-\frac{\Gamma}{2S+1}\sum_{i,j,\mu,\nu,\gamma,\gamma^{\prime}}c_{i\mu\gamma}^{\dagger}\left(1-\delta_{ij}\right)\left(1-\delta_{\mu\nu}\right)\nonumber \\
   & \quad\times\bra{\gamma}\left[\begin{aligned} & +\left(P_{i}^{0}\bm{\sigma}_{j}\cdot\bm{v}_{j}^{\prime}+\bm{\sigma}_{i}\cdot\bm{v}_{i}^{\prime}P_{j}^{0}\right)\\
   & -\frac{2iE_{g}}{\hbar\delta}\left(P_{i}^{0}\left(\Im\mel*{a_{1}}{\bm{\mathcal{B}}_{j}^{\left(l,s\right)}}{e}\cdot\bm{v}_{j}\right)\sigma_{0}+\left(\Im\mel*{a_{1}}{\bm{\mathcal{B}}_{i}^{\left(l,s\right)}}{e}\cdot\bm{v}_{i}\right)\sigma_{0}P_{j}^{0}\right)
  \end{aligned}
  \right]\ket{\gamma^{\prime}}c_{c_{j\nu\gamma^{\prime}}}\label{eq:3rd-orderHamiltonian}
\end{align}
where
\begin{align*}
\bm{\mathcal{B}}_{i}^{\left(l,s\right)} & =-\frac{\mu_{\text{B}}}{2c^{2}}\bm{E}_{i}^{\left(l,s\right)}\times\bm{r}_{i},
\end{align*}
and
\begin{align}
\bm{v}_{i}^{\prime} & =\frac{2i\lambda}{\hbar\delta} \mel{a_{1}}{\bm{L}_{i}}{e}\times\left(\Im\mel{a_{1}}{\bm{B}_{i}^{\left(l,s\right)}}{e}\times\bm{S}_{i}\right).\label{eq:3rd_v_term}
\end{align}
The vector \eqref{eq:3rd_v_term} corresponds to equation (12) in
the main text. Since the last term in equation \eqref{eq:3rd-orderHamiltonian}
does not related with the itinerant spin $\bm{\sigma}$, it does
not affect the localized spins at all. The terms including `vector potential' $-e\bm{A}_{i}^{\left(l,s\right)}$
in $\bm{\chi}_{i}^{\left(l,s\right)}$ vanish in the effective energy for the same reason. Besides, even within the secular approximation, the term with two optical vortex terms also exists as a perturbation. However, the magnitude of the perturbation term contributed by the optical vortex is small. In this study, we ignore the perturbation terms that contain more than two optical vortex absorption terms. Thus, we redefine $\mathcal{H}^{\left(3\right)}$
as
\begin{align*}
\mathcal{H}^{\left(3\right)} & =-\frac{\Gamma}{2S+1}\sum_{i,j,\mu,\nu,\gamma,\gamma^{\prime}}c_{i\mu\gamma}^{\dagger}\left(1-\delta_{ij}\right)\left(1-\delta_{\mu\nu}\right)\bra{\gamma}\left(P_{i}^{0}\bm{\sigma}_{j}\cdot\bm{v}_{j}^{\prime}+\bm{\sigma}_{i}\cdot\bm{v}_{i}^{\prime}P_{j}^{0}\right)\ket{\gamma^{\prime}}c_{j\mu\gamma^{\prime}}.
\end{align*}

\subsection{Effective spin-spin interaction Hamiltonian}

Defining the whole Hamiltonian $\mathcal{H}^{\prime}$ by summing
$\mathcal{H}^{\left(1\right)},\mathcal{H}^{\left(2\right)}$ and $\mathcal{H}^{\left(3\right)}$
for the total perturbation Hamiltonian, it becomes
\begin{align}
\mathcal{H}^{\prime} & =\mathcal{H}_{0}-\Gamma\sum_{i,j,\mu,\nu,\gamma,\gamma^{\prime}}c_{i\mu\gamma}^{\dagger}\left(1-\delta_{ij}\right)\left(1-\delta_{\mu\nu}\right)\bra{\gamma}P_{i}P_{j}\ket{\gamma^{\prime}}c_{j\nu\gamma^{\prime}},\label{eq:WholeHamiltonian}
\end{align}
where we define the projection operator as
\begin{align*}
P_{i} & =\frac{\left(\bm{S}_{i}+\bm{v}_{i}+\bm{v}_{i}^{\prime}\right)\cdot\bm{\sigma}+\left(S+1\right)\sigma_{0}}{2S+1}.
\end{align*}
The three mutually orthogonal vectors $\bm{S}_{i}$, $\bm{v}_{i}$, and $\bm{v}_{i}^{\prime}$ incorporate the effects of the first-, the second-, and the third-order perturbation, respectively. The Hamiltonian \eqref{eq:WholeHamiltonian} is just the one in equation (9) of the main text.

As is the derivation from \eqref{eq:ProjectionOperator0} to
\eqref{eq:FerromagneticHamiltonian}, we derive the effective Hamiltonian from $\mathcal{H}^{\prime}$ by substituting $\vartheta_{i}$ and $\varphi_{i}$ with $\vartheta_{i}+\Delta\vartheta_{i}$
and $\varphi_{i}+\Delta\varphi_{i}$ which are the angular coordinates
of the vector $\bm{S}_{i}+\bm{v}_{i}+\bm{v}_{i}^{\prime}$. The resultant spin-spin interactions become \cite{RN416}
\begin{align*}
\cos\left(\frac{\vartheta_{ij}+\Delta\vartheta_{ij}}{2}\right)\approx & J\left(\bm{S}_{i}+\bm{v}_{i}+\bm{v}_{i}^{\prime}\right)\cdot\left(\bm{S}_{j}+\bm{v}_{j}+\bm{v}_{j}^{\prime}\right)\\
\approx & J\bm{S}_{i}\cdot\bm{S}_{j} -\bm{D}_{ij}\cdot\left(\bm{S}_{i}\times\bm{S}_{j}\right)\\
 & +\bm{S}_{i}^{\dagger}\left(\bm{D}_{ij}\bm{\mathcal{B}}_{ij}^{\left(l,s\right)\dagger}+\bm{\mathcal{B}}_{ij}^{\left(l,s\right)}\bm{D}_{ij}^{\dagger}\right)\bm{S}_{j}.
\end{align*}
where
\begin{align*}
\bm{D}_{ij} & =-\frac{\lambda J}{E_{g}}\Im\left(\ev*{\bm{L}_{i}}_{\mu\nu}-\ev*{\bm{L}_{j}}_{\mu\nu}\right),\\
\bm{\mathcal{B}}_{ij}^{\left(l,s\right)} & =\frac{2E_{g}}{\hbar\delta}\Im\left(\ev*{\bm{\mathcal{B}}_{i}^{\left(l,s\right)}}_{\mu\nu}-\ev*{\bm{\mathcal{B}}_{j}^{\left(l,s\right)}}_{\mu\nu}\right).
\end{align*}
Here we set the magnitude of the Heisenberg coupling interaction as
$J$.

\section{Optical transition}

We show the theoretical form for the optical transition by the optical orbital angular momentum of the optical vortex. For the $3d$-orbitals of $\ce{Cr}$: $e,a_{1}$ and $e^{\prime}$
orbitals, we consider the interband transition of the $d$-electron
by optical vortices via the dipole interaction. The transition strength
from $Y_{2,0}$ to $Y_{2,m}$ ($m=\pm1,\pm2$) describes the following
integration,

\begin{align*}
 & \int drr^{2}\abs{R_{32}}^{2}\int d\theta d\varphi\sin\theta\begin{Bmatrix}Y_{2,+2}^{*}\\
Y_{2,-2}^{*}\\
Y_{2,+1}^{*}\\
Y_{2-1}^{*}
\end{Bmatrix}\bm{r}\cdot\left(\bm{\sigma}\times\bm{E}^{\left(l,s\right)}\right)Y_{2,0},
\end{align*}
where the radial function
\begin{align*}
R_{32}\left(r\right) & =\frac{4}{27\sqrt{10}}\left(\frac{1}{3r_{0}}\right)^{\frac{3}{2}}\left(\frac{r}{r_{0}}\right)^{2}e^{-\frac{r}{3r_{0}}},
\end{align*}
with the Bohr radius $r_{0}$ and spherical harmonics
\begin{align*}
Y_{2,0}\left(\theta,\phi\right) & =\sqrt{\frac{5}{16\pi}}\left(3\cos^{2}\theta-1\right),\\
Y_{2,\pm1}\left(\theta,\phi\right) & =\mp\sqrt{\frac{15}{8\pi}}\sin\theta\cos\theta e^{\pm i\phi},\\
Y_{2,\pm2}\left(\theta,\phi\right) & =\sqrt{\frac{15}{32\pi}}\sin^{2}\theta e^{\pm2i\phi}.
\end{align*}
For the transition strength, when $\left(l,s\right)=\left(+1,+1\right)$,
it becomes
\begin{align*}
  \int drr^{2}\abs{R_{32}}^{2}\int d\theta d\varphi\sin\theta Y_{2,+1}\bm{r}\cdot\left(\bm{\sigma}\times\bm{E}^{\left(l,s\right)}\right)Y_{2,0}
\approx  \frac{1}{4}\omega A_{0}r_{0}\dfrac{96}{\pi}\sqrt{\dfrac{3}{5}}\left(\dfrac{r_{0}}{w_{0}}\right)\ev*{\bm{\sigma}}_{\gamma\gamma^{\prime}}\cdot\begin{bmatrix}1\\
i\\
0
\end{bmatrix},
\end{align*}
\begin{align*}
  \int drr^{2}\abs{R_{32}}^{2}\int d\theta d\varphi\sin\theta Y_{2,+2}\bm{r}\cdot\left(\bm{\sigma}\times\bm{E}^{\left(l,s\right)}\right)Y_{2,0}
\approx  \omega A_{0}r_{0}\dfrac{96}{\pi}\sqrt{\dfrac{3}{5}}\left(\dfrac{r_{0}}{w_{0}}\right)\ev*{\bm{\sigma}}_{\gamma\gamma^{\prime}}\cdot\begin{bmatrix}0\\
0\\
-1
\end{bmatrix},
\end{align*}
\begin{align*}
\int drr^{2}\abs{R_{32}}^{2}\int d\theta d\varphi\sin\theta\begin{Bmatrix}Y_{2,-2}^{*}\\
Y_{2-1}^{*}
\end{Bmatrix}\bm{r}\cdot\left(\bm{\sigma}\times\bm{E}^{\left(l,s\right)}\right)Y_{2,0} & =0.
\end{align*}
On the other hand, when $\left(l,s\right)=\left(-1,-1\right)$, one can see
\begin{align*}
 \int drr^{2}\abs{R_{32}}^{2}\int d\theta d\varphi\sin\theta Y_{2,-1}\bm{r}\cdot\left(\bm{\sigma}\times\bm{E}^{\left(l,s\right)}\right)Y_{2,0}
\approx \frac{1}{4}\omega A_{0}r_{0}\dfrac{96}{\pi}\sqrt{\dfrac{3}{5}}\left(\dfrac{r_{0}}{w_{0}}\right)\ev*{\bm{\sigma}}_{\gamma\gamma^{\prime}}\cdot\begin{bmatrix}1\\
-i\\
0
\end{bmatrix},
\end{align*}
\begin{align*}
 \int drr^{2}\abs{R_{32}}^{2}\int d\theta d\varphi\sin\theta Y_{2,-2}\bm{r}\cdot\left(\bm{\sigma}\times\bm{E}^{\left(l,s\right)}\right)Y_{2,0}
\approx \omega A_{0}r_{0}\dfrac{96}{\pi}\sqrt{\dfrac{3}{5}}\left(\dfrac{r_{0}}{w_{0}}\right)\ev*{\bm{\sigma}}_{\gamma\gamma^{\prime}}\cdot\begin{bmatrix}0\\
0\\
1
\end{bmatrix},
\end{align*}
\begin{align*}
\int drr^{2}\abs{R_{32}}^{2}\int d\theta d\varphi\sin\theta\begin{Bmatrix}Y_{2,+2}^{*}\\
Y_{2+1}^{*}
\end{Bmatrix}\bm{r}\cdot\left(\bm{\sigma}\times\bm{E}^{\left(l,s\right)}\right)Y_{2,0} & =0.
\end{align*}
Even when $\left(l,s\right)=\left(+3,-1\right)$ or $\left(-3,+1\right)$,
the transition strength is still finite, but its order is the same
of or smaller than the one for $\left(l,s\right)=\left(\pm1,\pm1\right)$
multiplied by $\left(r_{0}/w_{0}\right)^{2}$.%

\providecommand{\newblock}{}